\documentclass[a4paper]{jpconf}
\usepackage{graphicx}
\usepackage{hyperref}

\begin{document}
\title{Event Index -- an LHCb Event Search System}
\author{A Ustyuzhanin \textsuperscript{1,2,6,7},
  A Artemov\textsuperscript{3,5}, N Kazeev \textsuperscript{1,2,4}
  and A Redkin\textsuperscript{3}}
\address{\textsuperscript{1} LHCb collaboration, Geneva, Switzerland}
\address{\textsuperscript{2} Yandex School of Data Analysis, Moscow, Russia}
\address{\textsuperscript{3} Yandex Data Factory, Moscow, Russia}
\address{\textsuperscript{4}
  Moscow Institute of Physics and Technology, Moscow, Russia}
\address{\textsuperscript{5} Moscow State University, Moscow, Russia}
\address{\textsuperscript{6} Kurchatov Institute, Moscow, Russia}
\address{\textsuperscript{7} National Research University Higher School of Economics (HSE)}
\ead{kazeevn@yandex-team.ru}

\begin{abstract}
  During LHC Run 1, the LHCb experiment recorded around $10^{11}$
  collision events. This paper describes Event Index --- an event search
  system. Its primary function is to quickly select subsets of events
  from a combination of conditions, such as the estimated decay
  channel or number of hits in a subdetector. Event Index is
  essentially Apache Lucene \cite{lucene} optimized for read-only
  indexes distributed over independent shards on independent nodes.
\end{abstract}

\section{Introduction}

The LHCb experiment records millions of proton collision events every
second. Most of them are not needed for further analysis and are
discarded by a sophisticated multi-layer trigger system
\cite{triggers}. What is left amounts to $10^{11}$ events in Run
1. Before physics analysis takes place, the number of events is
further reduced by a factor of around 10. This ``stripping'' process
takes place after the full reconstruction of the events, and produces
a set of a dozen ``streams'' of the analysis
dataset. \cite{LHCb-computing}. Those streams contain candidate events
for different processes --- identified by ``stripping lines.'' Events
that passed the stripping process are indexed by Event Index.

Along the stripping lines some other information is indexed --- global
activity counters (such as total number of tracks and hits in
individual subdetectors), logical file names (LFNs) on the GRID, and
run conditions database tags.

\section{Architecture}
\begin{figure}[!h]
  \centering
  \includegraphics[width=0.9\textwidth]{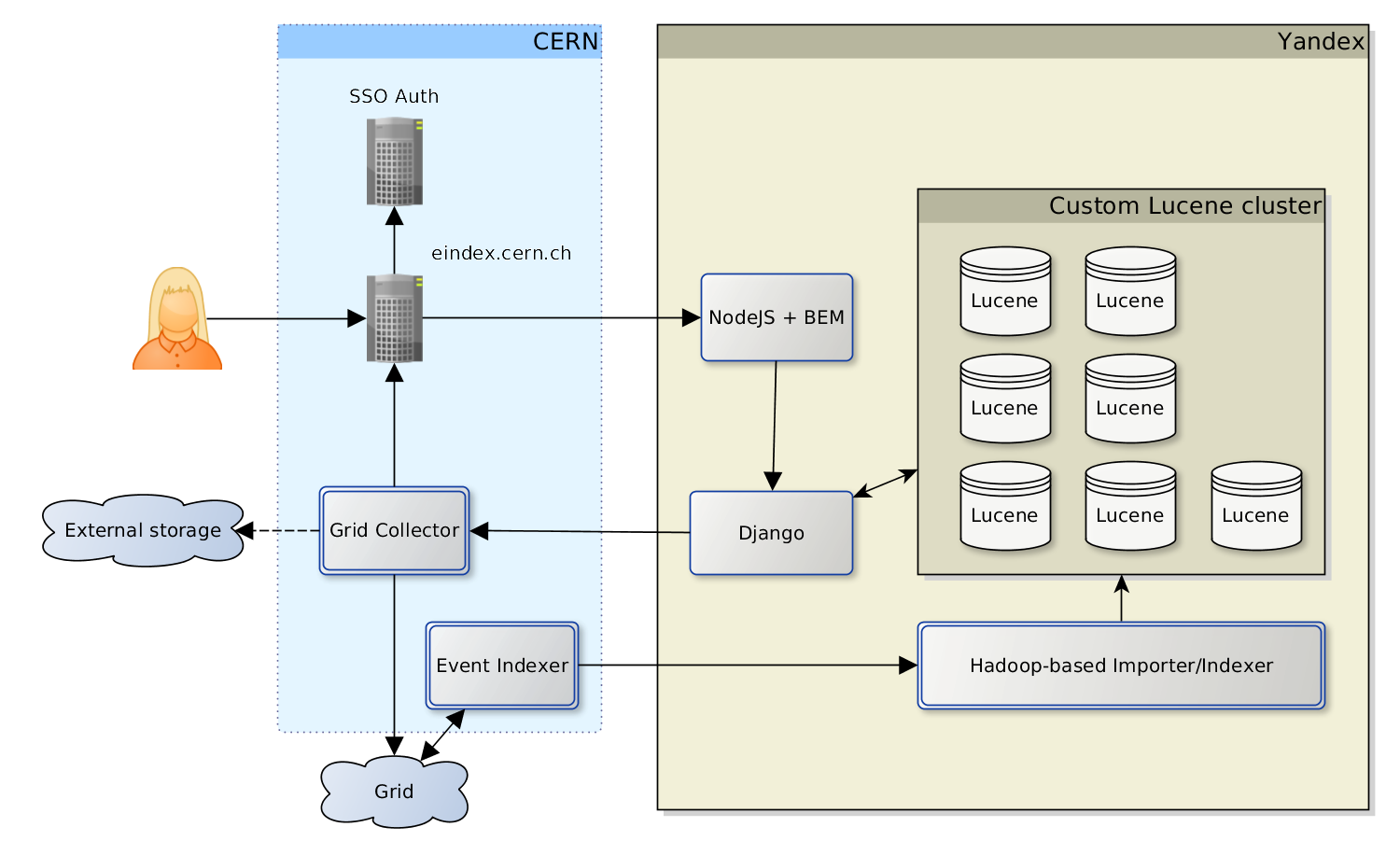}
  \caption{Event Index architecture}
  \label{fig:arch}
\end{figure}

Event Index consists of four primary parts: backend, which hosts the
indexes and processes the queries; frontend, which interacts with the
user; the GRID collector for downloading events from the GRID; and an
indexer for compiling the indexes. Their relationship is expressed on
the figure \ref{fig:arch}.

\subsection{Backend}
The principle component that stores events and handles queries is a
7-node cluster. Each node hosts several shards. A shard is an Apache
Lucene index. Indexes are build from .root files using MapReduce with
events being evenly distributed between the nodes.

Events are represented in backend in a problem-agnostic generic
format. Thus Event Index can be used on new datasets with minimal
modification.

Event Index is optimized for read-only indexes on a static hardware
configuration. Cluster expansion is still possible and can be
accomplished in two ways. First, if both new data and new nodes are
available, the data can be indexed on these nodes without changes to
the existing structure. This approach may be suboptimal, as the best
performance is only achieved when the data is evenly distributed among
the nodes. Second, if only nodes are added, we must either
redistribute the existing shards between nodes or reindex the dataset
to include them into the cluster. Index splitting is possible but
constitutes a highly experimental \cite{index-split} procedure with
computational costs similar to that of reindexing.

Requests are handled by a Java application as follows. Any node can
become a master node by virtue of initiating a request.

\begin{itemize}
\item Search request: A master node receives a query, sends it to all
  the nodes, each in turn sends it to its shards, shards run
  the query, and cache the resulting bitset.
\item Partial search results retrieval: A master node receives a
  query, asks all the nodes for the results counts, determines the
  nodes to send the request to. Nodes receiving the following request
  do the same with shards. The master node then gathers the responses
  and forwards them to the user.
\item Field value aggregation: A master node receives a query, sends it
  to all the nodes, each in turn sends it to its shards, each shard
  aggregates the field values from the matching events. The master node
  aggregates the results and returns them to the user.
\item Histogram calculation: A master node receives a query, sends it
  to all the nodes, each in turn sends it to its shards, each shard
  counts unique values of the requested fields, and returns them to
  the master node, which computes the resulting histogram.
\end{itemize}

Queries are transformed into Lucene Filters using a simple top-down
parser for context-free grammar. It consists of two parts: the
tokenizer and the parser itself. The tokenizer transforms a query
string into a list of tokens ($=, !=, >=, <=, (, ),$AND, OR, HAS) and
values. The parser uses the list to build the solution tree, using
prefix notation to handle parentheses and substituting HAS and AND for
missing unary and binary operators. It then converts the tree into a
Lucene Filter.

\subsection{Performance}

Indexing $10^{10}$ events took three days and 0.5 Tb of hard drive
space per node.

\begin{figure}[!h]
  \centering
  \includegraphics[width=6in]{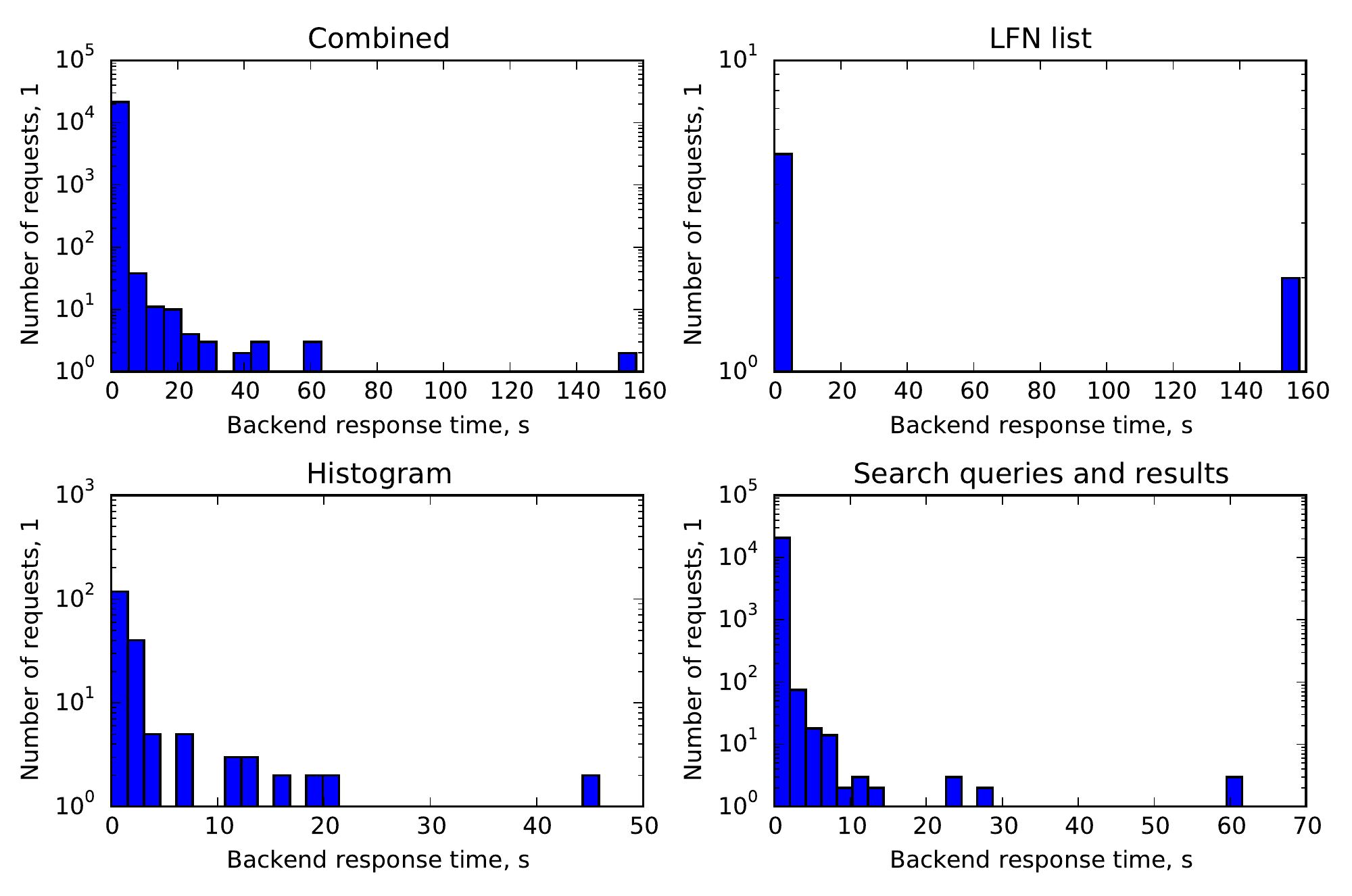}
  \caption{Backend response times for various request types. Data is
    taken from the live instance at https://eindex.cern.ch}
  \label{fig:performance}
\end{figure}

The backend response times for various requests can be seen in Figure
\ref{fig:performance}. This response was within 20 seconds for the
majority of requests. The outliers are currently being investigated.

\subsection{Frontend}
All user interaction is done through the web interface, protected by
CERN Single Sign-On \cite{sso}. Queries can either be typed manually
or constructed with the help of an interactive wizard. Example
searches:
\begin{itemize}
\item For a specific stripping line:
  \\ \mbox{``HAS
StrippingB02D0D0KSLLBeauty2CharmLineDecision AND AND stripping=20r1''}
\item By file location:
  \begin{sloppypar}
  ``lfn=LFN:/lhcb/LHCb/Collision11/CHARMTOBESWUM.DST/
     00022760/0002/00022760\_00029252\_1.CharmToBeSwum.dst
    AND stripping=20r1''
  \end{sloppypar}
\item Stripping line and nPVs value:
  \\ \mbox{``HAS
    StrippingB2D0KD2HHBeauty2CharmLineDecision AND stripping=21 AND
    nPVs$>4$''}
\end{itemize}

Event Index can compile a list of logical file names (LFN) containing
the search results.  If there are less than 1000 results, Event Index
can fetch them from GRID as a .root file and display them in the web
browser using Event Display \cite{EventDisplay}. Users can also plot
histograms for the global activity counters.

\subsection{The GRID collector}
The GRID collector handles the .root file download requests. It
resides on a dedicated server at CERN. It uses LHCb DIRAC
\cite{LHCbDIRAC} for retrieving event locations on the GRID. Then it
launches parallel Gaudi\cite{Gaudi} jobs for events retrieval and
format conversion for Event Display. The source code is available on
\url{https://gitlab.cern.ch/YSDA/grid_collector}.

\section{Status}
Event Index is deployed into production on
\url{https://eindex.cern.ch/} \footnote{accessible only to the members
  of LHCb collaboration}. Data from strippnigs 20, 20r1, 21, 21r1 is
available.

\section{Future works}
We are currently studying the needs of different groups in LHCb to
make Event Index a better tool. Plans include Python API, MC and turbo
stream \cite{turbo-stream} indexing, and free form query processing.

\section{Summary}
Event Index allows selection of events and viewing of histograms of
basic properties in a matter of seconds. This is much faster than the
current use of GRID, which can take hours. Event Index’s core
architecture will allow it to scale with data and be used for
different datasets.

\end{document}